\documentclass[10pt, conference]{IEEEtran}

\IEEEoverridecommandlockouts

\usepackage{cite}
\usepackage{amsmath,amssymb,amsfonts}
\usepackage{graphicx}
\usepackage{textcomp}
\usepackage{xcolor}

\usepackage{graphicx}
\usepackage{dcolumn}
\usepackage{bm}
\usepackage{booktabs}

\usepackage{algorithm}
\usepackage{algpseudocode}

\usepackage{orcidlink}
\usepackage[caption=false]{subfig}
\usepackage{qcircuit}
\usepackage{tabularx}
\usepackage{array}

\usepackage{braket}
\usepackage{url}
\usepackage{xurl} 

\begin{document}


\title{Spectral Gap Informed Ramp QAOA}

\author{
    \IEEEauthorblockN{
        Kieran McDowall\orcidlink{0009-0009-9074-9202}$^{1, *}$\thanks{$^{*}$Email: kieran.mcdowall@stfc.ac.uk}, 
        Konstantinos Georgopoulos\orcidlink{0000-0002-1481-2347}$^1$, and 
        Petros Wallden\orcidlink{0000-0002-0255-6542}$^2$
    }
    \IEEEauthorblockA{
        $^1$National Quantum Computing Centre, Rutherford Appleton Laboratory, \\
        Harwell Campus, Didcot, Oxfordshire, OX11 0QX, UK. 
    }
    \IEEEauthorblockA{
        $^2$Quantum Software Lab, School of Informatics, \\
        The University of Edinburgh, Edinburgh, UK.
    }
}

\maketitle

\begin{abstract}
 A challenge with the Quantum Approximate optimization Algorithm (QAOA), and variational algorithms in general, is finding good variational parameters, a task which in itself can be NP-hard. Recent work has sought to de-variationalise QAOA by picking well-informed guesses for the variational parameters. The Linear Ramp QAOA (LR--QAOA) \cite{montanez2025toward} achieves this by using parameter schedules inspired by the quantum adiabatic algorithm. 
 In this work, we propose Spectral Gap Informed Ramp QAOA (SGIR--QAOA), a new QAOA variant that incorporates spectral gap information from an adiabatic Hamiltonian, with the QAOA mixer Hamiltonian as the initial Hamiltonian, to construct smooth parameter schedules. SGIR--QAOA performs slow evolution where the spectral gap of the adiabatic Hamiltonian is small.
 We show that SGIR--QAOA has performance improvements over the LR--QAOA on Grover's problem at constant depth and that SGIR--QAOA requires shorter depths to achieve the same optimal solution probability. We then show that these performance benefits extend to a problem with potential practical applications -- the Maximum Independent Set (MIS) problem. Finally, we demonstrate the scalability of the SGIR--QAOA method using extrapolated spectral gap information for scales that the spectral gap cannot be exactly evaluated, and show that the advantage appears to persist under mild depolarising noise.
\end{abstract}


\maketitle

\section{Introduction}

Variational Quantum Algorithms (VQAs) have been extensively explored due to their suitability for running on NISQ devices. A well documented issue with VQAs is finding optimal variational parameters, a task which in itself can be NP-hard \cite{bittel2021training}. This issue arises from factors such as local minima and barren plateaus in the cost function landscape \cite{anschuetz2022quantum,mcclean2018barren}.

The Quantum Approximate optimization Algorithm (QAOA) \cite{farhi2014quantum} in the traditional sense can be viewed as a VQA. However, while VQAs in general can be viewed as heuristics with no provable performance guarantees, QAOA does have some theoretical proofs for speed-ups over classical algorithms while fixing its variational parameters \cite{boulebnane2024solving,montanaro2024quantum,farhi2025lower}. Adding to these theoretical proofs, other works have shown empirical evidence that QAOA may have a scaling advantage over classical solvers \cite{shaydulin2024evidence, montanez2025toward}.

To realise any potential quantum advantage that the QAOA may have, it is essential to address the variational parameters question. Studies have explored how the optimal parameters found when solving certain graph problems can be transferred to similar graph problems \cite{akshay2021parameter}. It is also possible to take a learning approach where one tries to find better generalised parameters during an expensive pre-training phase, applied to small dimensions, 
an approach taken 
in Ref. \cite{priestley2026practically}. Additionally, machine learning methods, particularly graph neural networks (GNNs), have been used to predict near-optimal QAOA parameters from graph structures \cite{khairy2020learning}.

Recently, Ref. \cite{montanez2025toward} introduced the Linear Ramp QAOA (LR--QAOA) where the QAOA parameters are taken to be simple linear schedules inspired by the Quantum Adiabatic Algorithm (QAA). However, the QAA with linear ramp schedules is known to be sub-optimal for certain problems; for instance, in Grover’s search, linear schedules fail to achieve the quadratic speed-up of Grover's algorithm \cite{roland2002quantum}. Optimal QAA schedules instead require slow evolution as the spectral gap of the adiabatic Hamiltonian closes. While other works have explored alternative QAOA parameter shapes \cite{cadavid2025bias,boulebnane2025quantum,kremenetski2021quantum}, these schedules are not constructed directly from the spectral gap. In particular, Ref. \cite{kremenetski2021quantum} investigated non-linear schedules for quantum chemistry, but these did not outperform linear ramps, suggesting that simply introducing non-linearity is insufficient. 

Our paper seeks to improve upon the LR--QAOA parameter schedule by using smooth Spectral Gap Informed Ramp QAOA (SGIR--QAOA) parameter schedules. These smooth parameter schedules are informed by the spectral gap of an adiabatic Hamiltonian that uses the QAOA mixer Hamiltonian as its initial starting point ($H_0$). An overview of the SGIR--QAOA methodology is provided in Figure \ref{plot:SGIR_fig1}.

Our main contributions are:

\begin{itemize}
    \item Showing empirically that the linear ramp schedule is not the optimal schedule for Grover's search problem. Our SGIR--QAOA achieves better performance in terms of optimal solution probability at constant depth $p$, and in terms of the depth $p$ required to reach a threshold optimal solution probability.
    \item We show that these advantages with SGIR--QAOA over LR--QAOA extend to the Maximum Independent Set (MIS) problem.
    \item We demonstrate the scalability of SGIR--QAOA by using a extrapolation technique for large problem sizes which incurs low overhead.
    \item We study performance under depolarising noise and show that noisy SGIR--QAOA can increase optimal solution probability in comparison to noisy LR--QAOA.
    \item We establish the regimes in which the advantage of SGIR--QAOA over LR--QAOA holds.
\end{itemize}


The paper is organised as follows: Section \ref{sect:preliminaries} contains the preliminaries necessary for understanding the SGIR--QAOA method including the LR--QAOA method (Section \ref{sect:preliminaries:LR--QAOA}), and the adiabatic Hamiltonian for Grover's problem (Section \ref{sect:preliminaries:Had}). Section \ref{sect:sgir} details the SGIR--QAOA method, the MIS problem is defined in Section \ref{sect:preliminaries:MIS}, Section \ref{sect:meth} explains the methodology used, Section \ref{sect:res} presents our results for both Grover's and the MIS problem, and in Section \ref{sect:disc} we discuss our findings.

\section{Preliminaries} \label{sect:preliminaries}

\subsection{Linear Ramp QAOA} \label{sect:preliminaries:LR--QAOA}


LR--QAOA is motivated by a linear QAA schedule and employs a linear parametrisation for the $2p$ QAOA variational parameters. While these parameters are not necessarily optimal, the primary advantage of LR--QAOA is that it reduces the optimization space to only two parameters: $\Delta_\beta, \Delta_\gamma$, which determine the gradients of the linear ramps. This reduction in variational parameters is especially advantageous at large QAOA depths $p$. The variational parameters for LR--QAOA are defined as:

\begin{equation}
    \beta_i= \left( 1-\frac{i}{p} \right) \Delta_\beta, \gamma_i = \frac{i+1}{p}\Delta_\gamma,
\end{equation}

with $i = 0,...,p-1$.

Different schedules for the QAA exist beyond the linear approach. For Grover's problem (the unstructured search problem with a single marked state), the optimal QAA schedule is smooth and evolves slowly near the minimum spectral gap of the adiabatic Hamiltonian, ensuring that the quadratic speed-up from Grover's algorithm is achieved \cite{roland2002quantum}. This speed-up can also be achieved using alternative schedules, such as in \cite{adamson2025adiabatic}. Conversely, if a linear QAA schedule is applied to Grover's problem, the quadratic speed-up is lost.

In the following sections, we introduce alternative parameter schedules: Roland Cerf -- QAOA (RC--QAOA), where the schedule used is from Ref. \cite{roland2002quantum}, where an analytical expression for the QAA schedule which achieves the quadratic speed-up on Grover's problem is derived; and Spectral Gap Informed Ramp -- QAOA (SGIR--QAOA) where the schedule is calculated from the eigenvalue spectrum of the adiabatic Hamiltonian with the QAOA mixer Hamiltonian as the initial Hamiltonian.

\subsection{An Adiabatic Hamiltonian for Grover's Problem} \label{sect:preliminaries:Had}

 The QAA involves interpolation from an initial, easy to prepare Hamiltonian, $H_0$, to a Hamiltonian whose ground state encodes the solution of a certain problem, $H_C$. If this evolution is performed slowly (or adiabatically) enough, the ground state is tracked throughout the evolution. The adiabatic Hamiltonian that describes this operation is:
\begin{equation} \label{eqn:Had}
    H_{Ad}=(1-t)H_0+tH_C.
\end{equation}

\begin{figure}[h]
\centering
\includegraphics[width=0.45\textwidth]{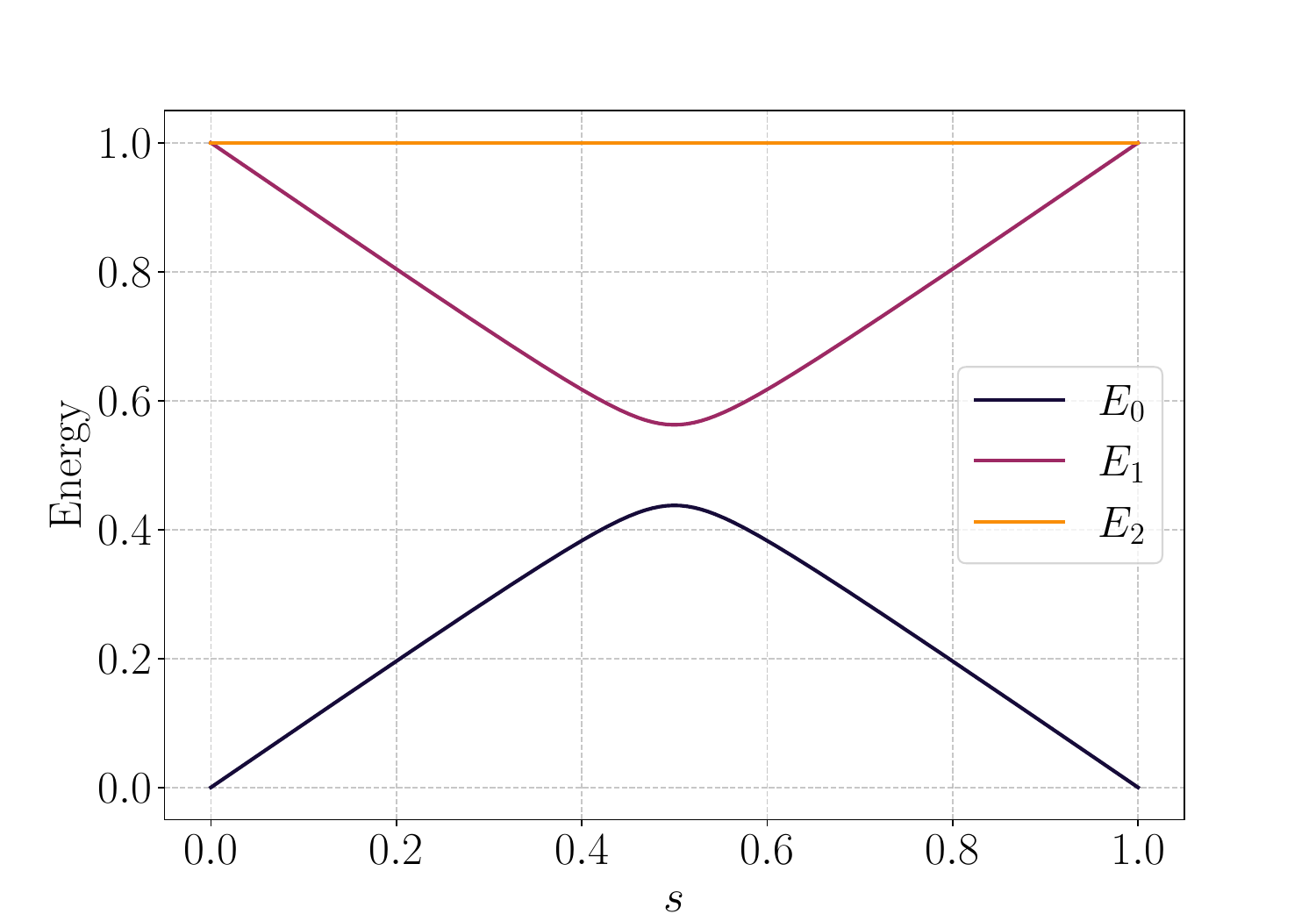}
\includegraphics[width=0.45\textwidth]{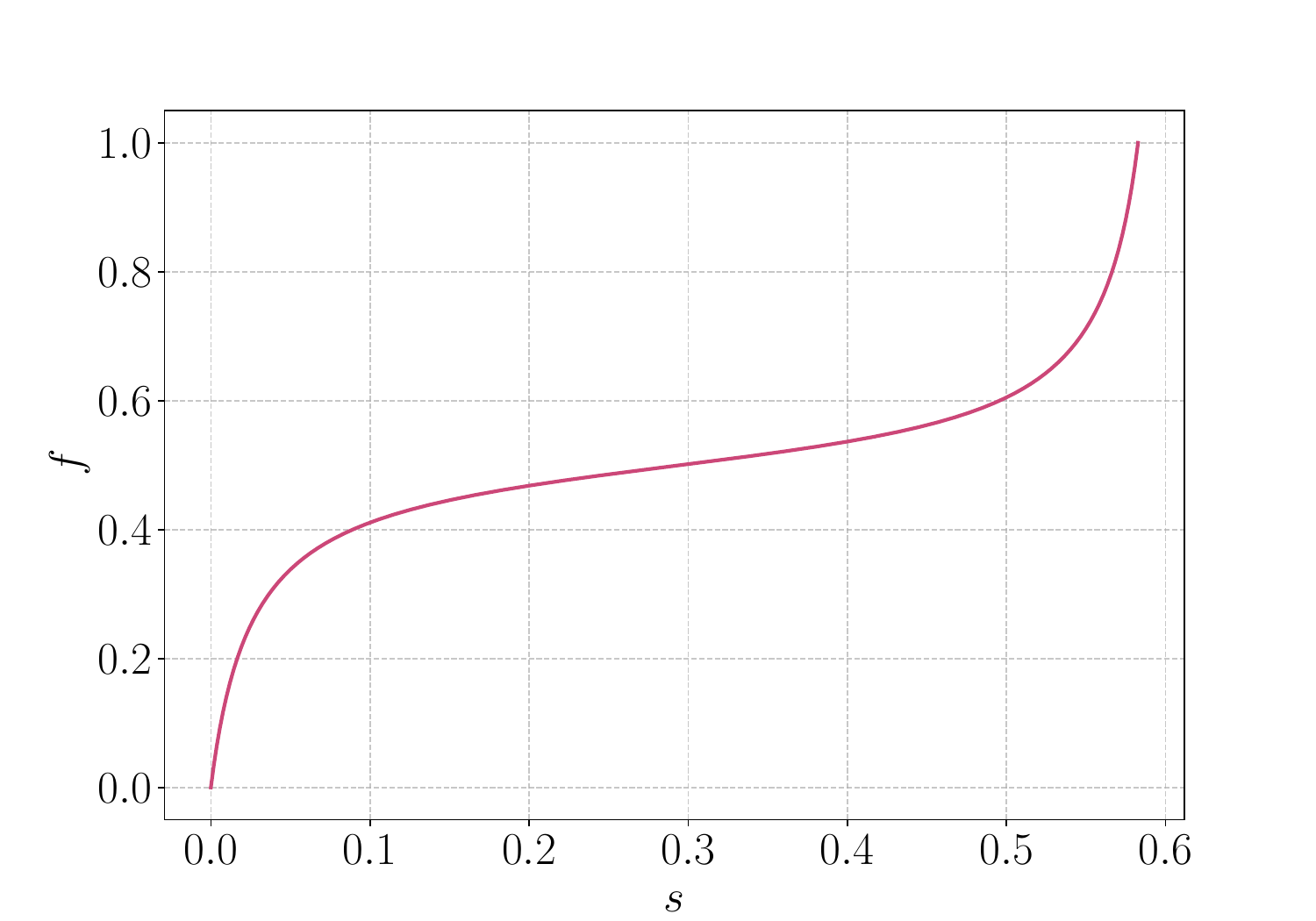}
\caption{\small{(Top) Eigenvalue spectrum from $H_{Ad}$ for Grover's problem with the marked solution state $|x\rangle=|000000\rangle$ $(n=6)$. (Bottom) The optimal Roland Cerf (RC) adiabatic schedule.}}
\label{plot:grover_spectrum_Had}
\end{figure}

For Grover's search problem we have:
\begin{equation}
    H_C = I - |\psi_{\text{sol}}\rangle\langle\psi_{\text{sol}}|,
\end{equation}
and
\begin{equation}
H_0 = I - |\psi_0\rangle \langle \psi_0|,
\end{equation}
where $|\psi_{\text{sol}}\rangle$ is the marked solution in the search problem and $|\psi_0\rangle = \frac{1}{\sqrt{N}} \sum_{i=1}^{N-1} |i\rangle.$ The eigenvalues for Grover's $H_{Ad}$ can be found analytically by realising that $H_{Ad}$ can be written as an invariant two-dimensional subspace, as in \cite{roland2002quantum,adamson2025adiabatic}. The analytical expression for the gap derived in \cite{roland2002quantum} is:
\begin{equation}
g(s) = \sqrt{1 - 4s(1 - s)\left(1 - \frac{1}{N}\right)} \, .
\end{equation} 
The optimal QAA schedule can then be found as its time derivative must be minimal as the gap is minimal: 
\begin{equation}
    \left|\frac{df}{ds}\right| \propto g^2(s),
\end{equation}
where $g(s)$ is the magnitude of the gap at all points $s$. An example of $H_{Ad}$ eigenvalues and its corresponding optimal schedule, $f(s)$, is shown in Figure \ref{plot:grover_spectrum_Had} for a particular instance of Grover's problem. We refer to using this schedule for QAOA parameters as Roland Cerf -- QAOA (RC--QAOA). However, it is vital to realise that an eigenvalue spectrum from an adiabatic Hamiltonian with a different initial Hamiltonian, $H_0$, takes a different shape. Choosing the initial adiabatic Hamiltonian to be the conventional QAOA mixer Hamiltonian is further explored in Section \ref{sect:sgir}.



\section{Spectral Gap Informed Ramp - QAOA} \label{sect:sgir}

\begin{figure*}[t]
\centering
\includegraphics[width=\textwidth]{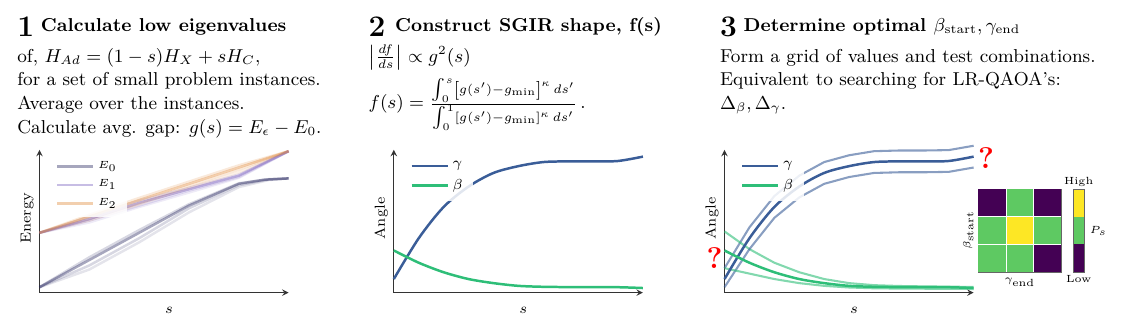}
\caption{\small{`Extrapolated' SGIR-QAOA methodology. \textbf{Step 1 -} The low energy eigenvalues of $H_{Ad}$ are calculated for varying s, where $s=[0,1]$ in steps $1/p$, where $p$ is the QAOA depth. For solving the MIS problem we average over 10 instances per problem size $n$, where small problem sizes are: $6 \leq n \leq 12$. The average gap $g(s)=E_\epsilon-E_0$ is then calculated, where $\epsilon$ depends on the degeneracy of the instance. This average $g(s)$ is then used for all instances $n>12$. The calculation to find these average gaps for the results in Figure \ref{plot:MIS_d3} took $5.8\,\mathrm{s}$. Note that in the `Exact' SGIR-QAOA version $g(s)$ is calculated for each instance. \textbf{Step 2 -} The SGIR shape, $f(s)$, is designed to evolve slowly when the gap is small. $f(s)$ is calculated using a weighted cumulative integral, where $\kappa$, is a hyperparameter and set to $\kappa=2$ in this work. \textbf{Step 3 -} The optimal $(\beta_\text{start}, \gamma_\text{end})$ are then found by forming a grid of possible values, then solving the particular problem instance with all the combination of values. The $(\beta_\text{start}, \gamma_\text{end})$ pair which yield the highest $P_s$ on the particular instance are selected.}}
\label{plot:SGIR_fig1}
\end{figure*}

Here we introduce the Spectral Gap Informed Ramp -- QAOA (SGIR--QAOA) method, summarised in Figure \ref{plot:SGIR_fig1}. Given that QAOA is a Trotterized discretisation of the QAA - applying alternating unitaries generated by a mixer Hamiltonian $H_{\text{mix}}$ and a cost Hamiltonian $H_C$ — we set the initial adiabatic Hamiltonian $H_0=H_{\text{mix}}$. With $H_0 = H_{\text{mix}} = H_X$, this gives us:

\begin{equation} \label{eqn:Hqaoa}
    H_{Ad}=(1-s)H_X+sH_C,
\end{equation}
 with $H_X = \sum_{i=1}^{n} X_i$. We sketch the eigenvalue spectrum for Equation \ref{eqn:Hqaoa} in Figure \ref{plot:SGIR_fig1} for a particular instance of Grover's problem along with the corresponding SGIR--QAOA schedule.


The eigenvalue spectrum in Figure \ref{plot:grover_spectrum_Had} differs from that given in Figure \ref{plot:SGIR_fig1}, due to the difference in $H_0$. We refer to the parameter schedule calculated from Equation \ref{eqn:Hqaoa} as a Spectral Gap Informed Ramp (SGIR). The variational parameters for SGIR--QAOA are defined as:

\begin{equation}
\begin{aligned}
\beta_i &=
\left(1-f\!\left(\frac{i}{p-1}\right)\right)\beta_{\mathrm{start}}
+
f\!\left(\frac{i}{p-1}\right)\beta_{\mathrm{end}}, \\
\\
\gamma_i &=
\left(1-f\!\left(\frac{i}{p-1}\right)\right)\gamma_{\mathrm{start}}
+
f\!\left(\frac{i}{p-1}\right)\gamma_{\mathrm{end}}.
\end{aligned}
\end{equation}

with $i = 0,\dots,p-1$, $\beta_\mathrm{start}, \gamma_\mathrm{end}$ parameters which must be searched for (equivalent to LR-QAOA's $\Delta_\beta, \Delta_\gamma$), $\beta_{\mathrm{end}}=\frac{1}{p}\beta_{\mathrm{start}},\gamma_{\mathrm{start}}=\frac{1}{p}\gamma_{\mathrm{end}}$, and $f(s)$ defined in Section \ref{sect:meth:sched_shape} Equation \ref{eqn:fs}. In the following section we define the MIS problem.

\section{Maximum Independent Set} \label{sect:preliminaries:MIS}

The Maximum Independent Set (MIS) problem is a combinatorial optimization problem where the goal is to find the largest subset of nodes in a graph such that none of these nodes share an edge. This problem has potential use-cases in sectors such as logistics and biology \cite{koch2025quantum}. 

The general MIS cost function can be written as:
\begin{equation}
C(\mathbf{x}) = \sum_{i \in V} x_i - \lambda \sum_{(i,j) \in E} x_i x_j,
\end{equation}
where $\lambda > 0$ is a penalty coefficient that enforces the independence constraint. The penalty term in our work is set to be excessively large, $\lambda=2000$. This is because we expect to see a greater separation in performance between LR--QAOA and SGIR--QAOA as the problem size increases, i.e., as the minimum spectral gap $g_{\text{min}}$ of the adiabatic Hamiltonian decreases. However, problem sizes where this occurs can lie outside the classically simuable region. Another way to reduce $g_{\text{min}}$ is by increasing the penalty term, $\lambda$, such that the problem sizes with performance separation now lie within the classically simuable region. In Appendix \ref{app:MIS}, Figure \ref{plot:changing_lamb} illustrates the effect of varying $\lambda$ on the results of our main experiment.

We study two types of MIS graphs: \emph{Erd\H{o}s--R\'enyi (ER) random graphs} with edge probability $p = 0.4$ (which we refer to here as dense graphs), and \emph{3-regular sparse graphs}, where each node has degree $3$. We found that the sparse degree 3 graphs were harder to solve with our QAOA methods, which is consistent with the fact that their maximum independent sets typically contain a larger number of nodes than in dense instances. To highlight the separation between LR--QAOA and SGIR--QAOA we chose to present the results on degree $3$ graphs in the main text while the dense ER graphs are included in Appendix \ref{app:MIS:dense}. 


\section{Methodology} \label{sect:meth}

Our experiments were performed with \textit{qiskit} \cite{javadi2024quantum}, using a state vector solver and $10,000$ shots.

\subsection{Eigenvalue Calculation} \label{sect:meth:eigs}

To plot the eigenvalue spectrum of a particular problem, eigenvalues of $H_{Ad}$ must be found at different times in the evolution, $s$. The discretisation is determined by the QAOA depth, $p$. However, finding the eigenvalues of a problem can be as difficult as solving the problem in the first place. To counter this we can use our extrapolation technique, which is further detailed in Section \ref{sect:meth:extrap}, to extend to large problem sizes. Using this extrapolation technique, it is only necessary to calculate the eigenvalue spectrum at small problem sizes.

Two methods are used to find the eigenvalues of $H_{Ad}$: a symmetry-reduced tridiagonal approach for Grover's problem and a matrix-free sparse approach for the MIS problem.

The symmetry-reduced approach exploits permutation symmetry in Grover's problem by restricting the dynamics to the symmetric subspace, whose basis states correspond to equal Hamming weight. This reduces the dimension of the Hamiltonian from $2^n$ to an $(n+1)$-dimensional tridiagonal matrix \cite{morley2019quantum}. The eigenvalues of this tridiagonal matrix are then computed using SciPy's \textit{eigh\_tridiagonal} routine \cite{2020SciPy-NMeth}. This approach is exact within the symmetry-reduced subspace and substantially cheaper than diagonalising the full $2^n \times 2^n$ Hamiltonian.

In the matrix-free approach, the Hamiltonian is never formed explicitly. Instead, eigenvalues are computed using a Krylov-subspace method (the implicitly restarted Lanczos algorithm via ARPACK), which accesses the Hamiltonian only through matrix-vector products $v \mapsto Hv$. Here, the \textit{eigsh} method from SciPy is used. This reduces memory usage and enables larger problem sizes than full dense diagonalisation. However, iterative sparse eigensolvers generally converge more reliably when low-lying eigenvalues are sufficiently separated, and degeneracies can make the identification of specific excited states less reliable.

\subsection{Schedule Calculation} \label{sect:meth:sched_shape}

To obtain the parameter schedule shape $f(s)$ from the calculated spectral gaps $g(s)$ a weighted cumulative integral is used. A monotonic mapping $f(s)$ is introduced such that
\begin{equation} \label{eqn:fs}
f(s) = \frac{\int_{0}^{s} \left[g(s') - g_{\min}\right]^\kappa \, ds'}{\int_{0}^{1} \left[g(s') - g_{\min}\right]^\kappa \, ds'} \, ,
\end{equation}
where $g_{\min} = \min_{s} g(s)$ ensures $\left|\frac{df}{ds}\right|=0$ at $g(s)=g_{\min}$, and $\kappa$ is a tunable exponent controlling the concentration of the schedule, set to  $\kappa=2$ in our work. For Grover's problem, $g_{\min}$ is set as the minimum gap between the ground state and the first excited state. For the MIS problem, a degeneracy check is carried out so that $g_{\min}$ is set as the minimum gap between the ground state and the next excited state that has energy distinct from the ground state. The resulting function $f(s)$ is normalised such that $f(0)=0$ and $f(1)=1$, and can be linearly rescaled to an arbitrary interval e.g. $[\gamma_\text{start}, \gamma_\text{end}]$.


\subsection{Finding the Optimal Schedule}

Once a schedule shape has been found for RC--QAOA and SGIR--QAOA from the eigenvalue spectrum, there is still the question: at what parameter values should these schedules start and end, $(\beta_\text{start}, \gamma_\text{end})$. This is equivalent to finding the $\Delta_\gamma,\Delta_\beta$ for LR--QAOA. To find these two parameters, for each QAOA schedule we perform a grid search, as done in Ref. \cite{dehn2026extrapolation}. For LR--QAOA, RC--QAOA and SGIR--QAOA we set the grid to be uniformly spaced in log-parameter space, with $\log(\beta_\text{start}), \log(\Delta_\beta) \in [-1.5, 0.5]$ and $\log(\gamma_\text{end}), \log(\Delta_\gamma) \in [-1, 1]$, discretised into an $11\times11$ grid.

Ref. \cite{dehn2026extrapolation} also shows how the search can be adjusted for larger problem sizes and how optimal $\Delta_\gamma,\Delta_\beta$ can be extrapolated. However, we use a constant grid across all problem sizes.

\subsection{Extrapolation} \label{sect:meth:extrap}

To extend the SGIR--QAOA method to large problem sizes without calculating the eigenvalue spectrum, an extrapolation technique can be used. For our MIS results in Section \ref{sect:res:MIS}, the extrapolated SGIR--QAOA schedule is obtained as follows: $g(s)$ for $n = 6-12$ is calculated. At $s=0$ we take the gap $E_2-E_0=g_2 = 4$, as we know this analytically. The minimum gap $g_\text{min}$ is found to always occur at $s=1$, we therefore use the approximation $g_{s=1}\approx0$. For $s$ between 0 and 1 we take the average gap values from $n = 6-12$ as there is minimal deviation across $n$. For the results presented in Figure \ref{plot:MIS_d3}, calculating the average gap profile $g(s)$ (by explicitly computing gaps for $n = 6-12$ at $p=10$) takes $5.8\,\mathrm{s}$. The execution time of this step increases with both the range of problem sizes evaluated and the circuit depth $p$. Preliminary empirical testing suggests that this step displays sub-quadratic runtime scaling with $p$. The overheads of the SGIR--QAOA and LR--QAOA methods are summarised in Table \ref{tab:resource_comparison}.

\begin{algorithm}
\caption{Extrapolated SGIR--QAOA for MIS}\label{alg:sgir-extrapolated}
\begin{algorithmic}[1]
\State Compute an average gap profile $\bar g(s)$ from small MIS instances ($6\leq n\leq 12$).
\State Set $\bar g(0)=4$ and $\bar g(1)=0$.
\State Construct $f(s)$ via the weighted cumulative integral in Eq.~\eqref{eqn:fs}.
\For{each $(\beta_{\mathrm{start}},\gamma_{\mathrm{end}})$ on the log-grid}
    \State Set $\beta_{\mathrm{end}}=\beta_{\mathrm{start}}/p$ and $\gamma_{\mathrm{start}}=\gamma_{\mathrm{end}}/p$.
    \For{$i=0$ to $p-1$}
        \State $s_i=i/(p-1)$ and define $\beta_i,\gamma_i$ from $f(s_i)$.
    \EndFor
    \State Run QAOA, post-select feasible MIS bitstrings, and compute $P_s$.
\EndFor
\State Choose the parameters with maximal $P_s$.
\end{algorithmic}
\end{algorithm}

In Section \ref{sect:res} results are shown for both extrapolated SGIR--QAOA and exact SGIR--QAOA. Here, exact SGIR--QAOA refers to calculating the eigenvalue spectrum at all problem sizes. We distinguish the two SGIR methods in our figure legends. 



\subsection{Evaluation metrics} \label{sect:meth:eval_metrics}

In this section we define our evaluation metrics and justify the plots that we use to assess the performance of our different methods. 

\textbf{Optimal Solution Probability,} $\boldsymbol{P_s}$ -- The proportion of times, $N_{\text{ground state}}$, the ground state occurs in all $N_{\text{solutions}}$ solutions:
\begin{equation*}
P_s = \frac{N_{\text{ground state}}}{N_{\text{solutions}}}.
\end{equation*}

The optimal solution (ground state) is found using the classical Gurobi solver.

$P_s$ is plotted at varying problem size $n$ for a constant QAOA depth $p$. It is important here to note that the exponential scaling behaviour of QAOA at constant depth can change for larger problem sizes. To suppress diabatic transitions as problem size increases the required $p$ to maintain approximate adiabacity must increase.

For this reason we include another plot which is the required QAOA depth $p$ to reach a threshold optimal solution probability $P_s^{th}$. A threshold is arbitrarily set, in practice $P_s^{th}=\frac{1}{n_{\text{shots}}}$ is adequate so as to obtain the optimal solution once. For the purpose of our study we pick a $P_s^{th}$ which allows us to show obvious separation between our methods at the explored problem sizes. At each problem size $n$, a QAOA depth $p$ is tested over $10$ instances and an average $P_s^{\text{avg}}$ is calculated. If $P_s^{\text{avg}}\geq P_s^{th}$ then then the depth $p$ is recorded and we move on to the next problem size. If  $P_s^{\text{avg}}< P_s^{th}$ then $p$ is increased and the process is repeated.

Our QAOA depth scaling with problem size plots are inspired by \cite{dehn2026extrapolation} where the growth in depth of LR--QAOA is related to its runtime. 

Another metric which can be used to quantify the quality of solution is the approximation ratio which describes the proximity of the energy expectation value to the true minimum energy value \cite{mcdowall2025practical}. At the problem sizes tested in our study we are able to obtain optimal solutions, therefore, we chose to focus on the optimal solution probability. However, in Appendix \ref{app:MIS} Figure \ref{plot:MIS_d3_AR} we include a plot of the approximation ratio for the same experiment as in Figure \ref{plot:MIS_d3}.


\subsection{Post-Selection} \label{sect:meth:post-sel}

For the MIS results presented, a post-selection routine is used to remove infeasible solutions. These removed solutions are solutions that do not form an independent set. The probabilities are not renormalised for $P_s$ so as to show raw, true success probability.

\section{Results} \label{sect:res}

\subsection{Grover's Problem} \label{sect:res:grovers}

As an initial test of our method we solve Grover's problem, as the eigenvalue spectrum for its adiabatic Hamiltonian is known analytically \cite{roland2002quantum,adamson2025adiabatic,morley2019quantum}. This allows us to test whether using the eigenvalue spectrum from Equation \ref{eqn:Hqaoa} is appropriate, and we do this by comparing SGIR--QAOA to RC--QAOA. It is important to note that for combinatorial optimization problems in general an analytical expression for its eigenvalues does not exist. We solve the MIS problem in Section \ref{sect:res:MIS}, where we demonstrate an extrapolation technique which only uses the eigenvalue spectrum at small problem sizes.

\begin{figure}[h]
\centering
\includegraphics[width=0.5\textwidth]{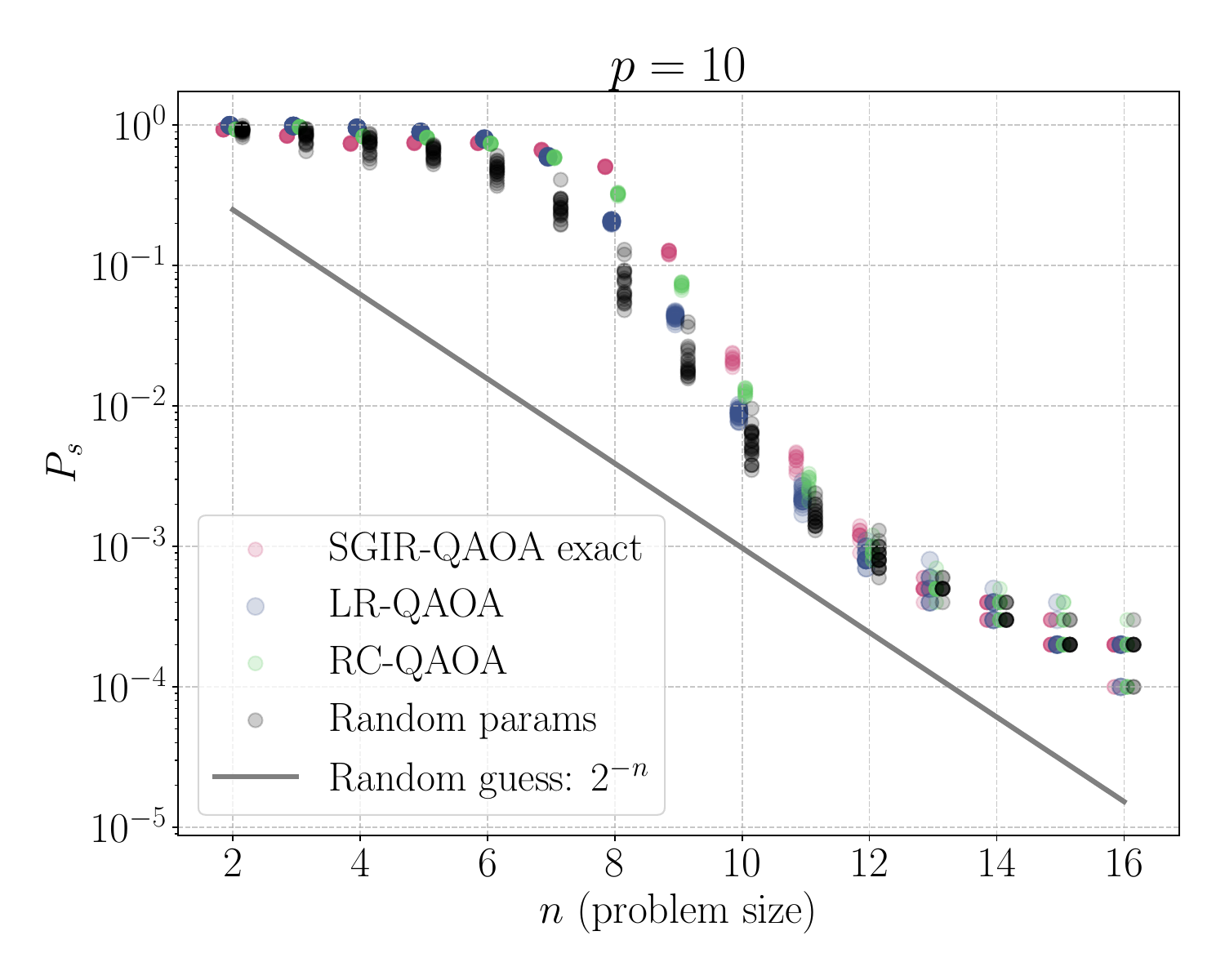}
\caption{\small{Solving Grover's problem with our QAOA methods at constant QAOA depth $p=10$. At each $n$ the experiment is repeated 10 times with a different random marked solution. The random QAOA parameters are changed for each of these instances. The y-axis is logarithmic.}}
\label{plot:grover_wSGIR}
\end{figure}

In Figure \ref{plot:grover_wSGIR}, we solve instances of Grover's problem for varying size $n$ with QAOA $p=10$. The marked solution is randomized 10 times for each $n$. SGIR--QAOA is compared against LR--QAOA, RC--QAOA and QAOA with random parameters, where the parameters are selected at random for each problem instance. For $n\leq5$, we observe that all methods perform approximately equally well, with random parameters providing a lower bound on performance. This is attributed to the relative ease of the problem, resulting from the large spectral gap compared with the QAOA depth $p=10$. As $n$ increases, all methods undergo a phase transition in terms of the exponential scaling of their optimal solution probability, $P_s$. Between $6 \leq n \leq 12$ we see SGIR--QAOA returning the highest $P_s$. For these problem sizes the minimum spectral gap $g_{\text{min}}$ has decreased, where slower evolution at $g_{\text{min}}$ is increasingly vital to avoid diabatic transitions. The correlation between the performance improvement of SGIR--QAOA over LR--QAOA with respect to $g_{\text{min}}$ is captured in Figure \ref{plot:percent_improve_gap_and_n}.

 This behaviour persists until $n=13$, where all methods converge to similar performance. In this region, the QAOA depth $p$ is now insufficient for any of our QAOA techniques, where performance is now independent of the parameter schedule. Both the onset of this independence and the point of maximum performance divergence between SGIR--QAOA and LR--QAOA are dependent on the chosen depth $p$.

 

\begin{figure}[h]
\centering
\includegraphics[width=0.5\textwidth]{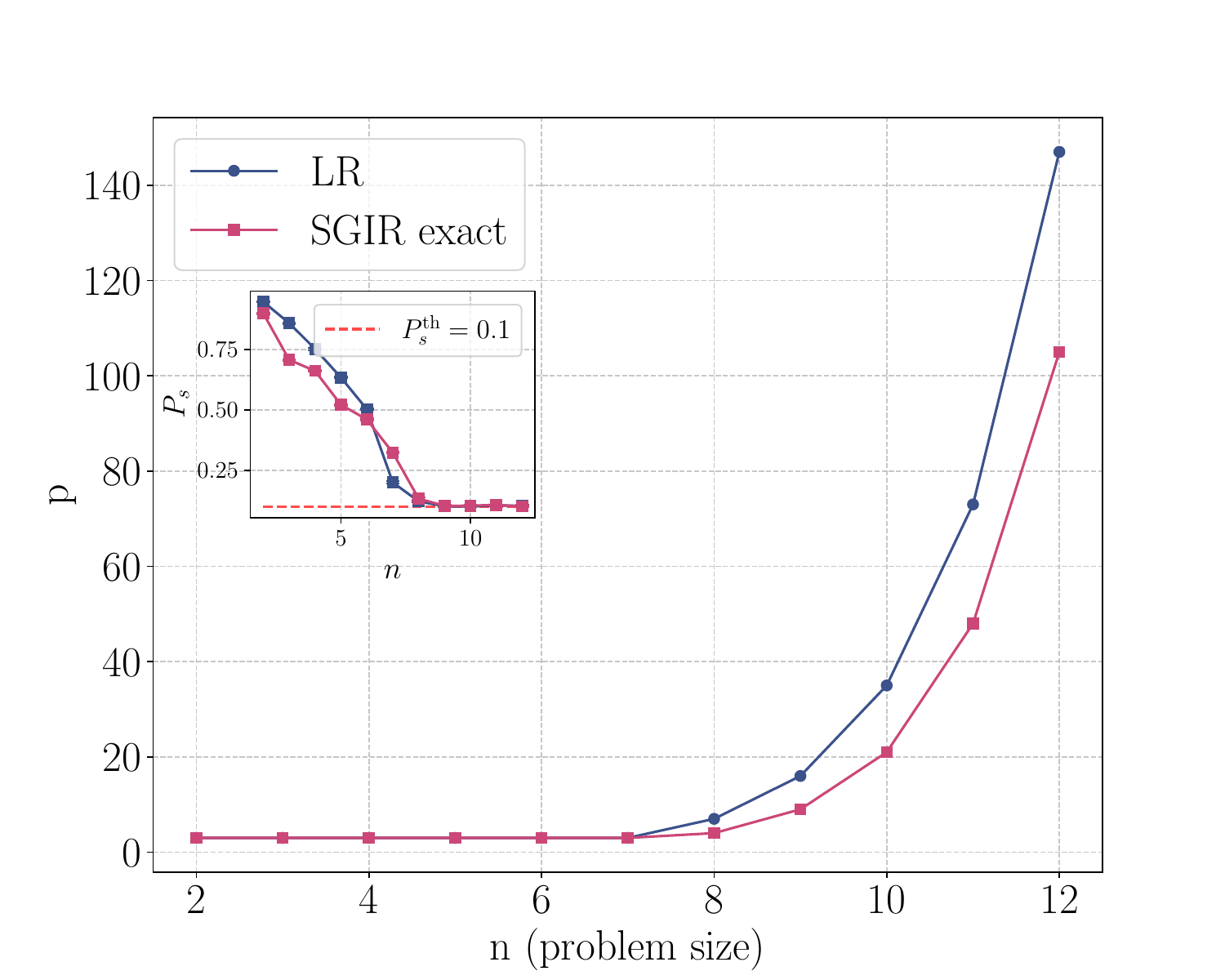}
\caption{\small{Solving Grover's problem where the QAOA depth $p$ required to reach an optimal solution threshold $P_s^{th}$ at different problem sizes is plotted. }
}
\label{plot:grover_p_with_n}
\end{figure}

We further explore the performance of SGIR--QAOA in Figure \ref{plot:grover_p_with_n} where the required QAOA depth $p$ to reach a threshold probability $P_s^{th}=0.1$ is plotted against the problem size for LR--QAOA and SGIR--QAOA.

It can be seen from Figure \ref{plot:grover_p_with_n} that the required QAOA depth is smaller for SGIR--QAOA than LR--QAOA past $n=7$. As problem size increases the difference in the required $p$ for SGIR--QAOA and LR--QAOA increases. This is a powerful result as when running on noisy hardware deeper circuits result in larger noise accumulation. Performance under noise is further explored in Section \ref{sect:res:MIS:noise}.

\subsection{Maximum Independent Set} \label{sect:res:MIS}

\begin{figure}[h]
\centering
\includegraphics[width=0.5\textwidth]{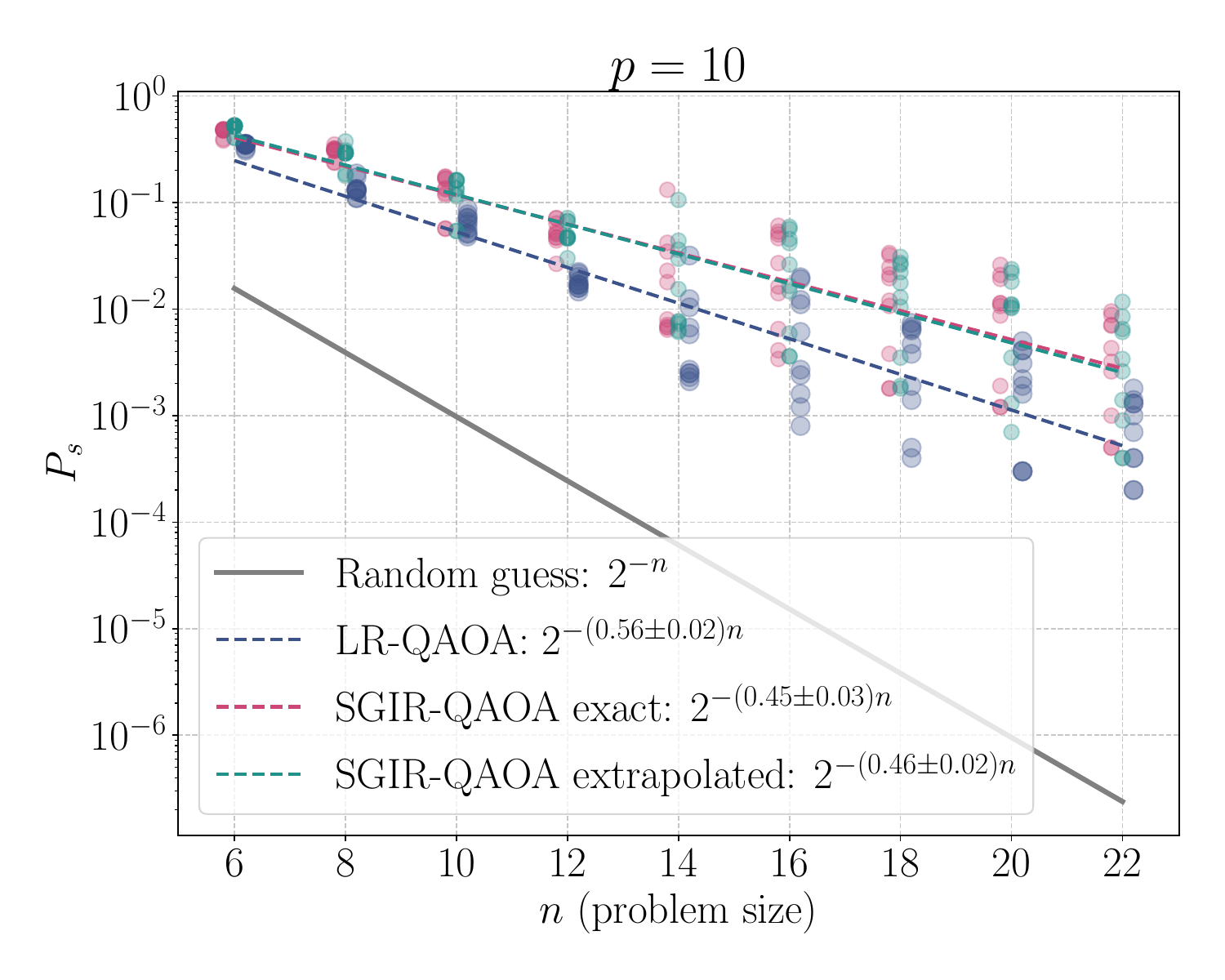}
\caption{\small{
Solving MIS degree 3 graphs with LR--QAOA, exact SGIR--QAOA and extrapolated SGIR--QAOA. For extrapolated SGIR--QAOA, the $n = 6-12$ gaps are calculated which inform the SGIR--QAOA schedule for $n>12$. At each $n$ the experiment is repeated 10 times with a different randomly generated instance. A log scale is used on the y-axis. The error in the fit for the exponential scaling coefficient is included in the legend.
}}
\label{plot:MIS_d3}
\end{figure}

In Figure \ref{plot:MIS_d3} we plot the achieved $P_s$ when solving random instances of degree 3 MIS problems at varying problem size with QAOA depth $p=10$. The exponential scaling achieved with exact SGIR--QAOA is $2^{-(0.45\pm0.03)n}$ compared with $2^{-(0.56\pm0.02)n}$ for LR--QAOA.

For the extrapolated SGIR--QAOA in Figure \ref{plot:MIS_d3}, we still observe a significant performance advantage over LR--QAOA for the tested problem sizes: $2^{-(0.46\pm0.02)n}$ vs $2^{-(0.56\pm0.02)n}$. In Appendix \ref{app:MIS}, Figure \ref{plot:MIS_extrap_d3_large_n} investigates solving at larger problem sizes. 

\begin{figure}[h]
\centering
\includegraphics[width=0.5\textwidth]{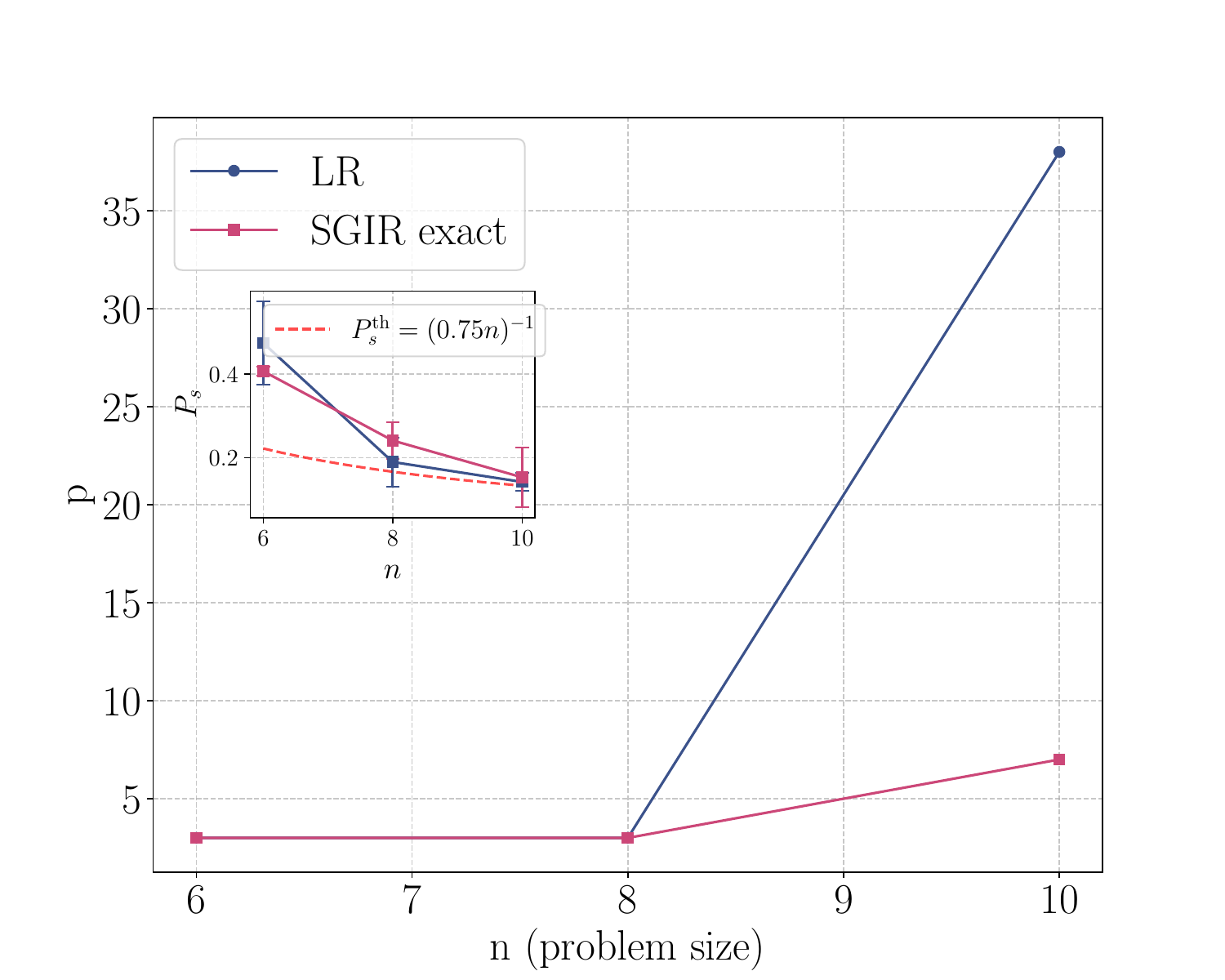} 
\caption{\small{
Solving the MIS problem with degree 3 graphs where the QAOA depth $p$ required to reach an optimal solution threshold $P_s^{th}$ at different problem sizes is plotted.}}
\label{plot:MIS_p_with_n}
\end{figure}

In Figure \ref{plot:MIS_p_with_n} we show the required QAOA depth $p$ required to reach a threshold optimal solution probability at varying problem size. Here, the threshold is set to $P_s^{th}=\frac{1}{0.75n}$. It was found that a constant $P_s^{th}$ required extremely large QAOA depth $p$, hence, the relaxed threshold used here. Even with the threshold set at $P_s^{th}=\frac{1}{0.75n}$ problem sizes above $n=10$ were infeasible to obtain solutions for, due to the large $p$ required.

\subsubsection{Maximum Independent Set with Noise} \label{sect:res:MIS:noise} 
\begin{figure}[h]
\centering
\includegraphics[width=0.5\textwidth]{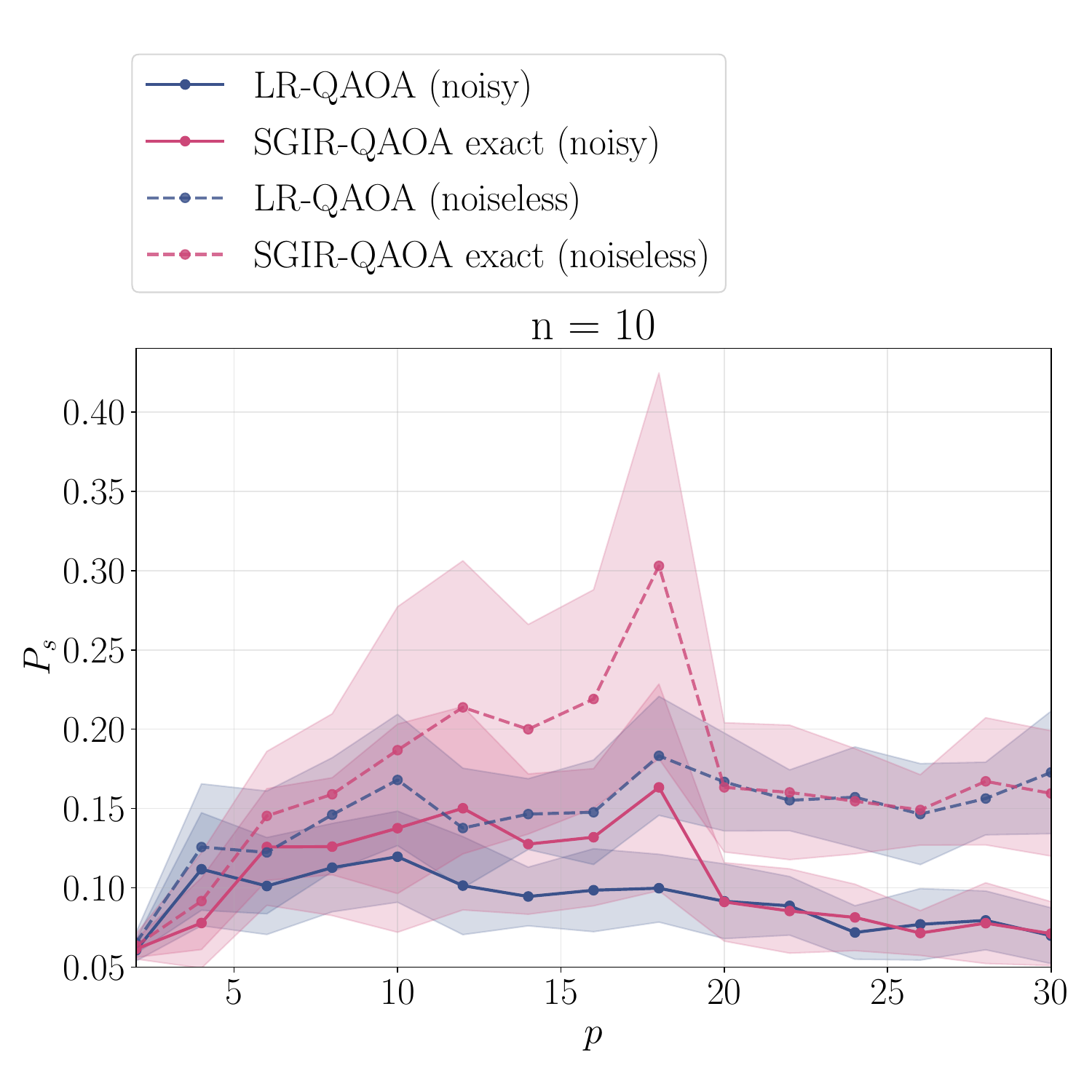} 
\caption{\small{
Solving the MIS problem (degree three graphs) under depolarising noise. Averaging over 10 instances at the 10 node problem size. $\lambda = 100$, searching $20 \times 20$ $(\beta_\text{start}, \gamma_\text{end})$ and $(\Delta_\gamma,\Delta_\beta)$ values for each $p$. The shaded region correspond to the standard deviation error of the noisy points. In Figure \ref{plot:large_p_behav} we plot the noiseless LR--QAOA and SGIR--QAOA showing behaviour at larger $p$ .
}}
\label{plot:MIS_noise_n10}
\end{figure}

~\\
In this section the effect of depolarising noise on the performance of SGIR--QAOA and LR--QAOA is explored. In Figure \ref{plot:MIS_noise_n10}, 10 random $n=10$ MIS instances are solved at varying QAOA depth $p$ under depolarising noise of strength $p_{\text{noise}}=0.001$ and with noiseless statevector simulation as a comparison, with the penalty term now set at $\lambda = 100$ and the parameter grid set to $20\times20$. Under noise, the optimal solution probability for SGIR--QAOA is higher than that of LR--QAOA up to $p=20$. In Figure \ref{plot:large_p_behav} we plot the noiseless LR--QAOA and SGIR--QAOA at larger values of depth $p$.

\section{Discussion} \label{sect:disc}

SGIR--QAOA can achieve performance improvements over LR--QAOA in terms of both optimal solution probability at constant depth and the required QAOA depth, $p$, to reach an optimal solution probability threshold. 


These performance improvements for SGIR--QAOA exist for Grover's problem and extend to an optimization problem with practical interest -- the MIS problem. The good performance of SGIR--QAOA should also hold for the MIS compliment problems, namely, the maximum clique problem and the minimum vertex problem, and we expect that performance improvements for SGIR--QAOA also hold for combinatorial optimization problems in general. For this to be shown, the problem must be difficult enough in terms of its minimum spectral gap and cost landscape. Performance separation, therefore, might lie outwith the classically simulable range for certain `easy' problems. For any QUBO problem with penalty terms it should be possible to show performance separation by increasing the penalty to term to a large value to emulate what happens at large problem sizes.

SGIR--QAOA provides advantages over LR--QAOA under two further conditions:
\begin{itemize}
    \item \textit{A limited depth $p$ budget}: If the depth $p$ is sufficiently large for a relatively small problem with a large spectral gap, performance is independent of the parameter schedule. This behaviour is illustrated in Figure \ref{plot:grover_wSGIR}, $n<6$.
    \item \textit{Sufficient depth $p$ relative to problem size $n$}: Conversely, SGIR--QAOA loses its advantage at large problem sizes $n$ (characterised by a small spectral gap) if the depth $p$ is insufficient. In this regime, performance is once again independent of the chosen parameter schedule, as demonstrated in Figure \ref{plot:grover_wSGIR}, $n>12$.
\end{itemize}

The scalability of the SGIR--QAOA method is addressed by demonstrating that the improvements in exponential scaling over LR--QAOA at constant depth hold when using a low-overhead extrapolation technique. Although this extrapolation technique is problem specific, the methodology used here can be transferred to different combinatorial optimization problems. Furthermore, the performance improvements of SGIR--QAOA are maintained under depolarising noise.

Among the additional areas that could be explored in future work are comparisons with standard QAOA and other parameter transfer approaches. Of particular interest would be a comparison of both solution quality and the computational resources required to obtain the optimal parameters for each method.

Beyond comparisons with other QAOA techniques, SGIR--QAOA could also be benchmarked against classical solvers. In Ref. \cite{montanez2025toward}, results suggest that LR--QAOA may provide a scaling advantage for hard instances of the weighted Max-Cut problem. SGIR--QAOA may potentially improve upon this scaling behaviour; however, it is important to note that these results are obtained at constant $p$, and that the scaling behaviour will change once $p$ is no longer sufficient for the problem size. A more meaningful metric for QAOA runtime scaling is therefore not the time-to-solution estimated from the inverse of the optimal solution probability at fixed depth, but rather the QAOA depth required to maintain a constant success probability, or even a success probability that decreases with problem size (such as $1/n$). Demonstrating reductions in the depth required to achieve a target probability will therefore be essential for establishing any potential scaling advantages of QAOA.

Finally, it would be valuable to connect SGIR--QAOA with recent efforts to better understand the mechanisms underlying QAOA performance. In particular, equivalences between QAOA and quantum annealing have been established \cite{boulebnane2025quantum}, while analyses of QAOA dynamics have highlighted connections between the eigenstates of the cost and mixer Hamiltonians \cite{kremenetski2023quantum}. Exploring whether SGIR--QAOA alters these eigenstate interactions may provide insight into the origins of its improved performance compared with LR--QAOA.

\noindent \textbf{Note added regarding concurrent work:} After the initial version of this paper, we became aware of related independent work by Ref. \cite{nzongani2026scaling}. Their approach employs a training phase to learn average schedules from random QUBO instances, which are then applied to the Max-Cut problem and compared against optimal QAOA parameters. Our work focuses on Grover's search and the MIS problem, providing comparisons with alternative QAOA schedules.





\section*{Acknowledgements}

The authors would like to thank Sebastian Brandhofer for his discussions on LR--QAOA, MIS, and quantum optimization in general. The authors would also like to thank Stephen DiAdamo and Qoro Quantum for help with MPS simulation using the Maestro simulator \cite{bertomeu2025maestro}. PW was supported by EPSRC grants EP/X026167/1, EP/Z53318X/1 and the QATCH Programme.

\section*{Code Availability}

The code used to generate the experimental results presented can be found in the public GitHub repository \cite{kmREPO}.

\bibliographystyle{unsrt} 
\bibliography{mybib}

\appendix

\subsection{Resourcing}

In Table~\ref{tab:resource_comparison} we summarise the computational resources required by the extrapolated SGIR--QAOA and compare these with those of LR--QAOA.

\begin{table*}[htbp]
\centering
\caption{Comparison of computational resources required across each step for LR--QAOA and Extrapolated SGIR--QAOA.}
\label{tab:resource_comparison}
\begin{tabular}{p{3.4cm} p{4.2cm} p{4.2cm} p{4.5cm}}
\toprule
\textbf{Method} & \textbf{Step 1: Gap Profile $g(s)$} & \textbf{Step 2: Schedule $f(s)$} & \textbf{Step 3: Parameter Search} \\
\midrule
\textbf{LR--QAOA} & 
\textbf{None} \newline (No spectral analysis required) & 
\textbf{None} \newline (Linear ramp schedule is pre-defined) & 
\textbf{2D Grid Search \newline(or classical optimisation)} \newline Search for parameters $(\Delta_\beta, \Delta_\gamma)$. \\
\addlinespace
\textbf{Extrapolated SGIR--QAOA} & 
\textbf{Low Classical Pre-computation} \newline Computes ground and low excited states for small instances. Overhead depends on the range of instances used to calculate the average $g(s)$, and the depth $p$ (e.g., for MIS $n = 6\text{--}12$, $p=10$, this takes $5.8\,\mathrm{s}$). Provisional evidence of sub-quadratic scaling with $p$. & 
\textbf{Negligible $\approx0.2\,\mathrm{ms}$} \newline Fast evaluation of Eqn. \ref{eqn:fs} which is constant with $n$ and $p$. & 
\textbf{2D Grid Search \newline(or classical optimisation)} \newline Search for parameters $(\beta_{\text{start}}, \gamma_{\text{end}})$. \\
\bottomrule
\end{tabular}
\end{table*}

\subsection{SGIR--QAOA Performance with $g_{\text{min}}$} \label{app:grovers:perf_expl}

\begin{figure}[h]
\centering
\includegraphics[width=0.5\textwidth]{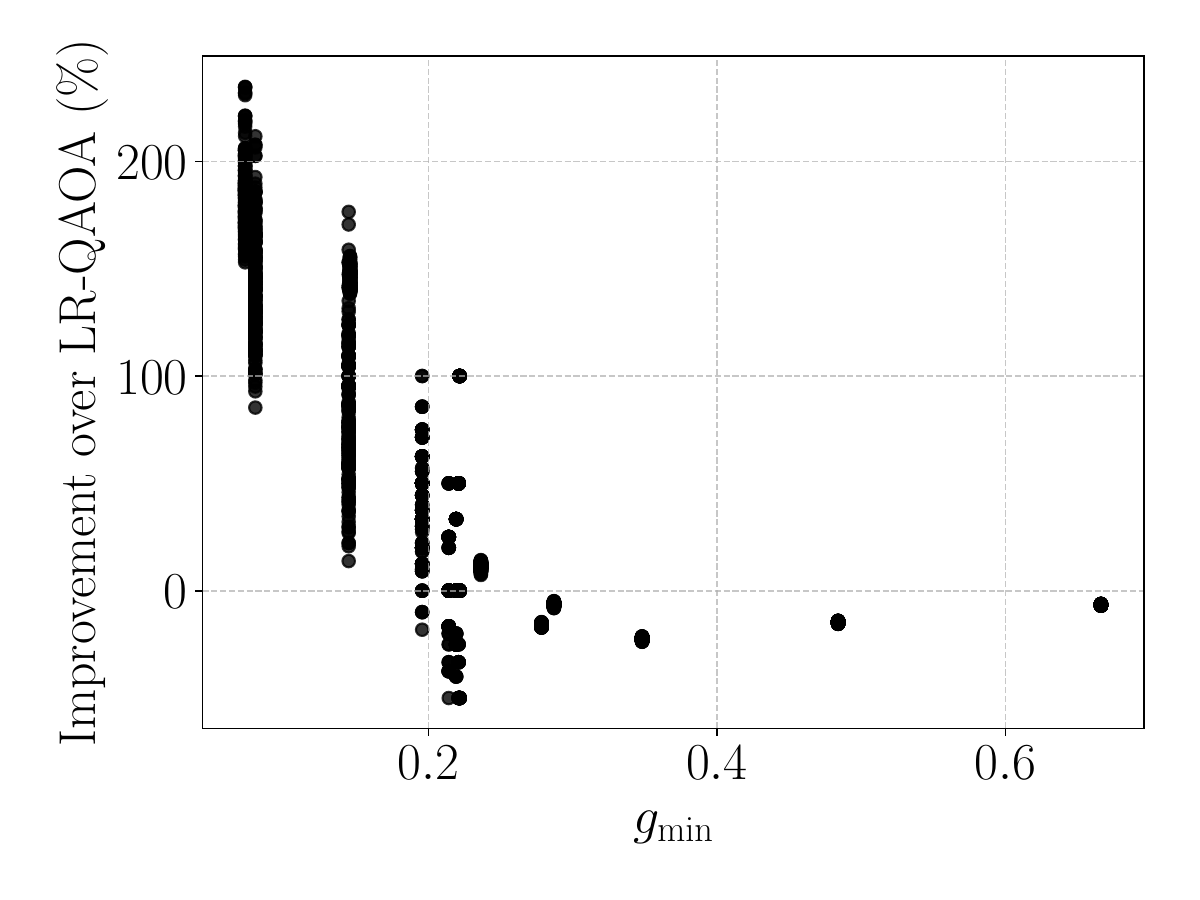}
\caption{\small{The percentage improvement of SGIR--QAOA against LR--QAOA, $p=10$, with the minimum spectral gap solving Grover's problem. The Pearson correlation coefficient is $\rho = -0.68$.}}
\label{plot:percent_improve_gap_and_n}
\end{figure}

 In Figure \ref{plot:percent_improve_gap_and_n} the correlation between the performance improvement of SGIR--QAOA over LR--QAOA for Grover's problem with respect to $g_{\text{min}}$ is shown.

\subsection{Maximum Independent Set - Extras} \label{app:MIS}

\begin{figure}[!htbp]
\centering
\includegraphics[width=0.5\textwidth]{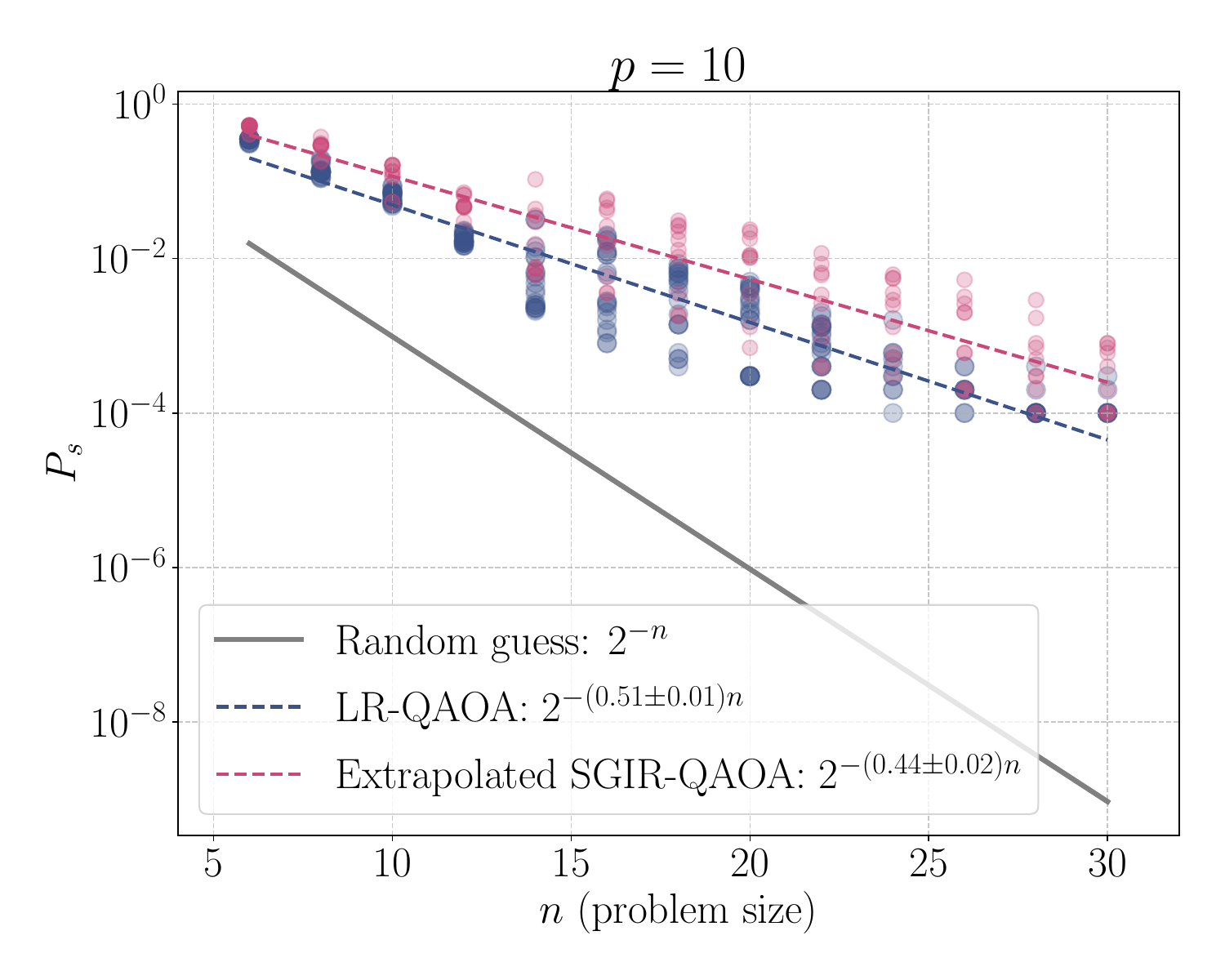}
\caption{\small{
Solving MIS degree 3 graphs with extrapolated SGIR--QAOA at larger n than in Figure \ref{plot:MIS_d3}. The gaps for $n = 6-12$ are calculated which inform the SGIR--QAOA schedule for $n>12$. At each $n$ the experiment is repeated 10 times with a different randomly generated instance. A log scale is used on the y-axis. The error in the fit for the exponential scaling coefficient is included in the legend.
}}
\label{plot:MIS_extrap_d3_large_n}
\end{figure}

In Figure \ref{plot:MIS_extrap_d3_large_n} we explore the behaviour of extrapolated SGIR--QAOA and LR--QAOA at larger values of $n$. The separation in Figure \ref{plot:MIS_extrap_d3_large_n} between the exponential scaling of SGIR--QAOA and LR--QAOA has fallen at these larger sizes compared with the scaling calculated from Figure \ref{plot:MIS_d3}. This is due to $p=10$ becoming insufficient at these large problem sizes. The performance of the two methods therefore begins to converge (like in Figure \ref{plot:grover_wSGIR}) at the larger $n$.

\begin{figure}[h]
\centering
\includegraphics[width=0.5\textwidth]{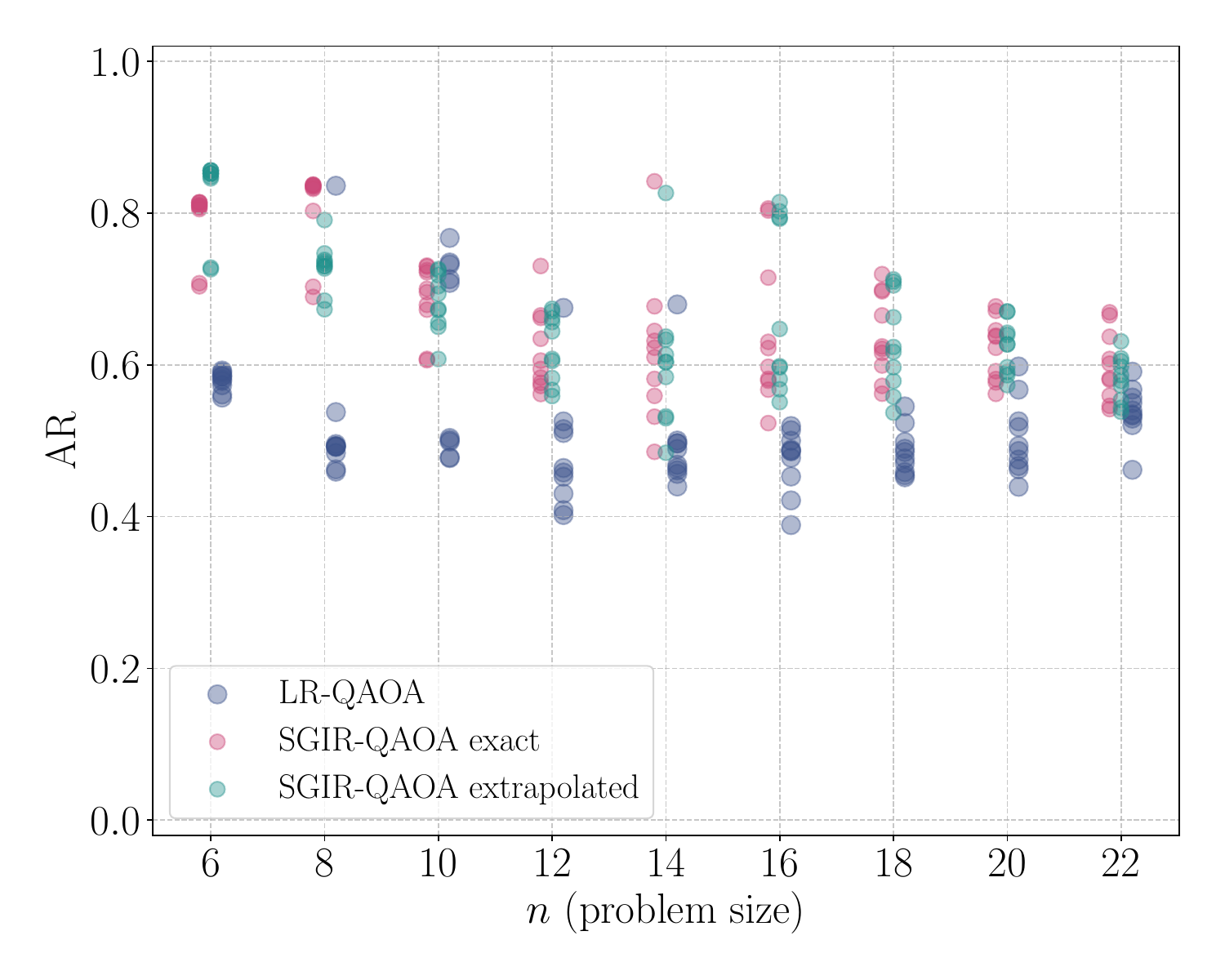}
\caption{\small{
Solving MIS degree 3 graphs with LR--QAOA, exact SGIR--QAOA and extrapolated SGIR--QAOA. For extrapolated SGIR--QAOA, the $n = 6-12$ gaps are calculated which inform the SGIR--QAOA schedule for $n>12$. At each $n$ the experiment is repeated 10 times with a different randomly generated instance. A log scale is used on the y-axis. The error in the fit for the exponential scaling coefficient is included in the legend. The approximation ratio is calculated post-selection. 
}}
\label{plot:MIS_d3_AR}
\end{figure}

In Figure \ref{plot:MIS_d3_AR}, we present the approximation ratio (AR) for the same experiment shown in Figure \ref{plot:MIS_d3}. The AR is defined as $AR = \frac{E_\text{avg}}{E_\text{min}}$, where the optimal value $E_\text{min}$ is calculated using the Gurobi solver. Note that $E_\text{avg}$ is computed over post-selected, valid independent sets. Consequently, the probabilities of these feasible states must be renormalised to sum to unity in order to yield a mathematically valid expectation value.

\begin{figure}[h]
\centering
\includegraphics[width=0.5\textwidth]{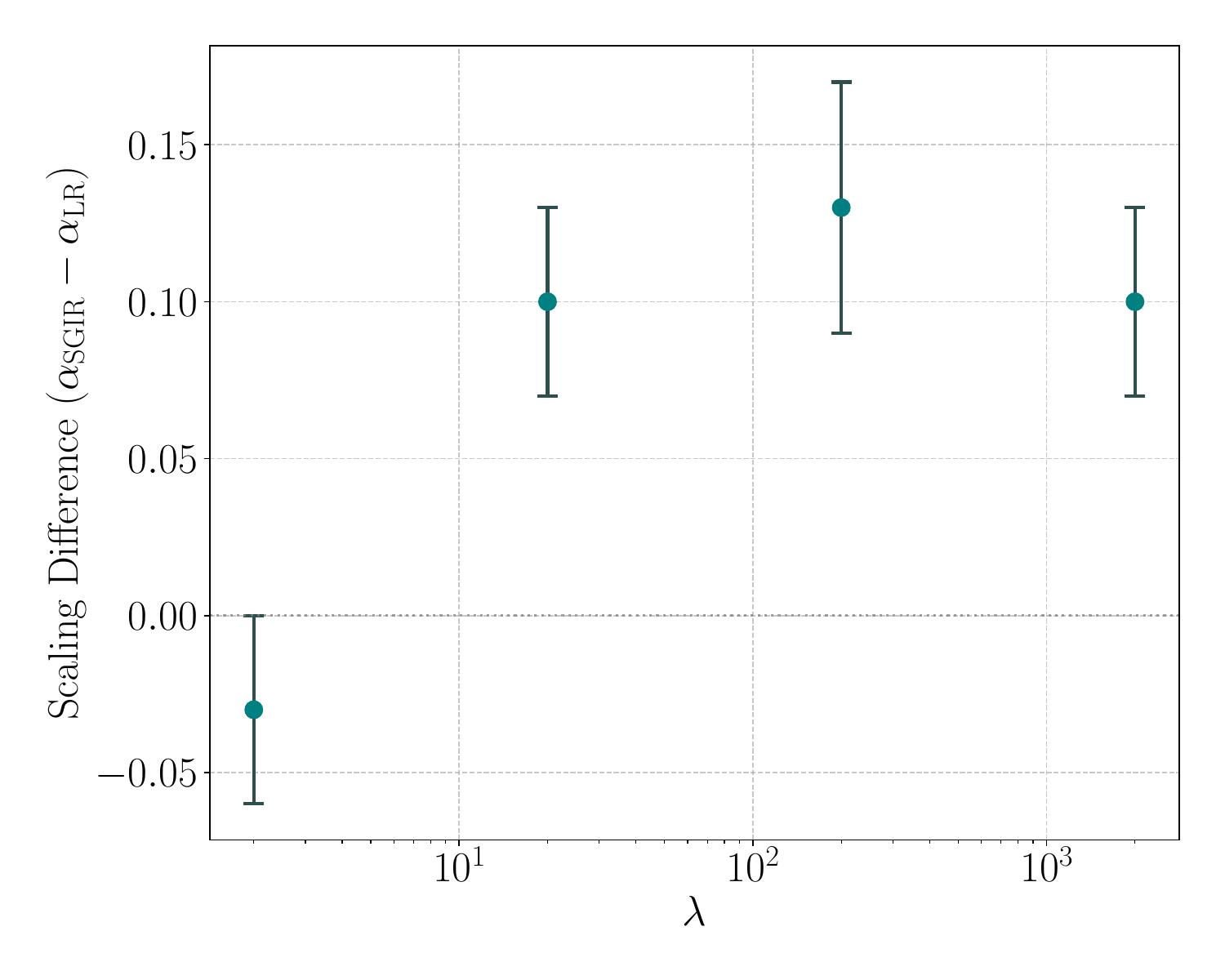}
\caption{\small{Solving MIS degree 3 graphs with LR--QAOA and extrapolated SGIR--QAOA for varying penalty term $\lambda$. For each $\lambda$, a plot like Figure \ref{plot:MIS_d3} is produced, and the difference in the scaling coefficeint $\alpha$ between SGIR extrapolated and LR--QAOA is calculated. A log scale is used on the x-axis. Error bars show the combined uncertainty in the scaling difference which comes from the uncertainty in the fit for the exponential scaling coefficient.
}}
\label{plot:changing_lamb}
\end{figure}

In Figure \ref{plot:changing_lamb}, we investigate the effect of varying the penalty term $\lambda$, using the same experimental setup as in Figure \ref{plot:MIS_d3}.

\begin{figure}[!htbp]
\centering
\includegraphics[width=0.5\textwidth]{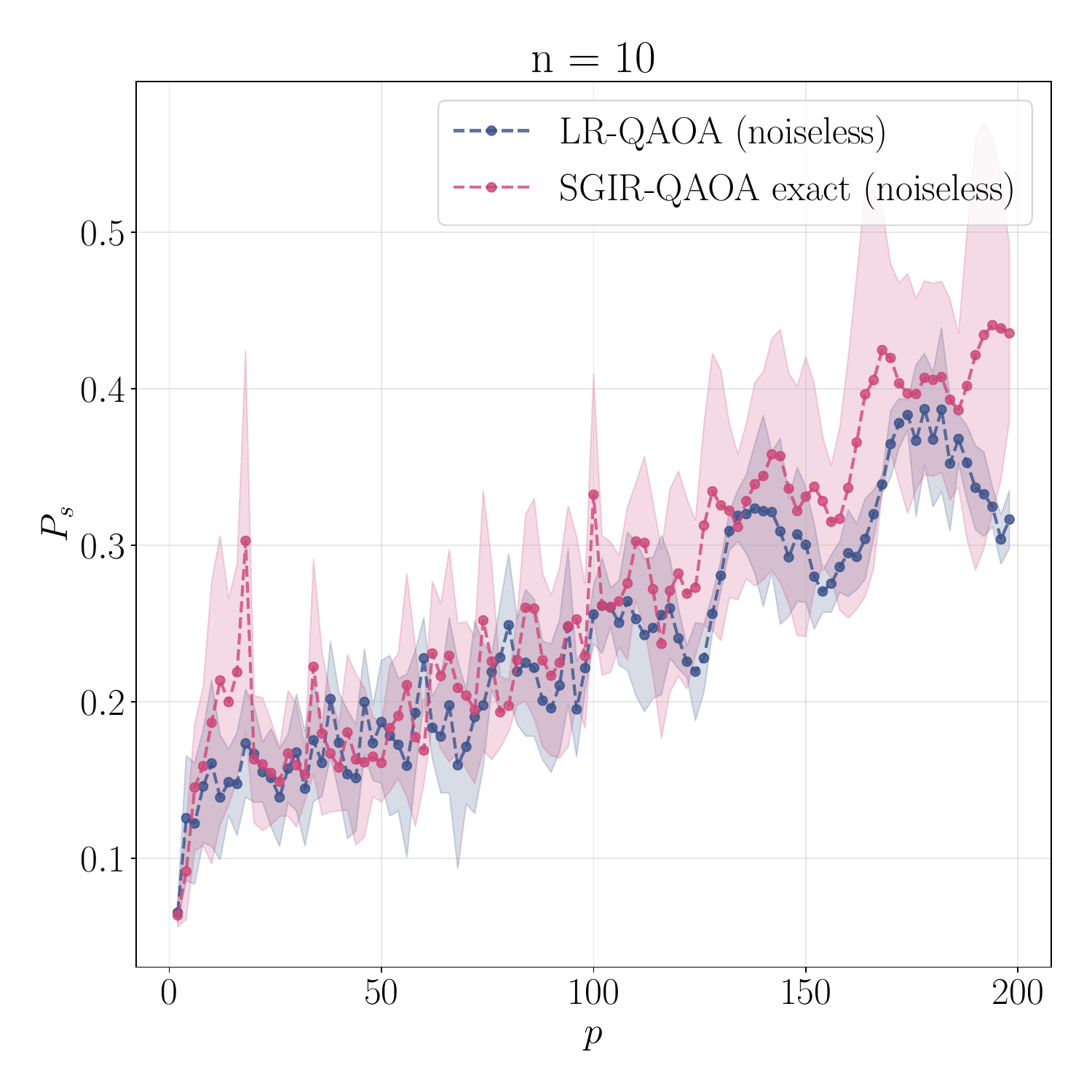} 
\caption{\small{
An extension of Figure \ref{plot:MIS_noise_n10}.
Solving the MIS problem (degree three graphs) with a noiseless statevector solver at larger $p$ values. Averaging over 10 instances at the 10 node problem size. $\lambda = 100$, searching $20 \times 20$ $(\beta_\text{start}, \gamma_\text{end})$ and $(\Delta_\gamma,\Delta_\beta)$ values for each $n$. The shaded region correspond to the standard deviation error.
}}
\label{plot:large_p_behav}
\end{figure}

In Figure \ref{plot:large_p_behav} we plot the same experiment as in Figure \ref{plot:MIS_noise_n10} but now we extend the noiseless LR--QAOA and SGIR--QAOA to larger $p$ values. 


\subsubsection{Dense Erd\H{o}s--R\'enyi (ER) Random Graphs} \label{app:MIS:dense} 

\begin{figure}[!htbp]
\centering
\includegraphics[width=0.5\textwidth]{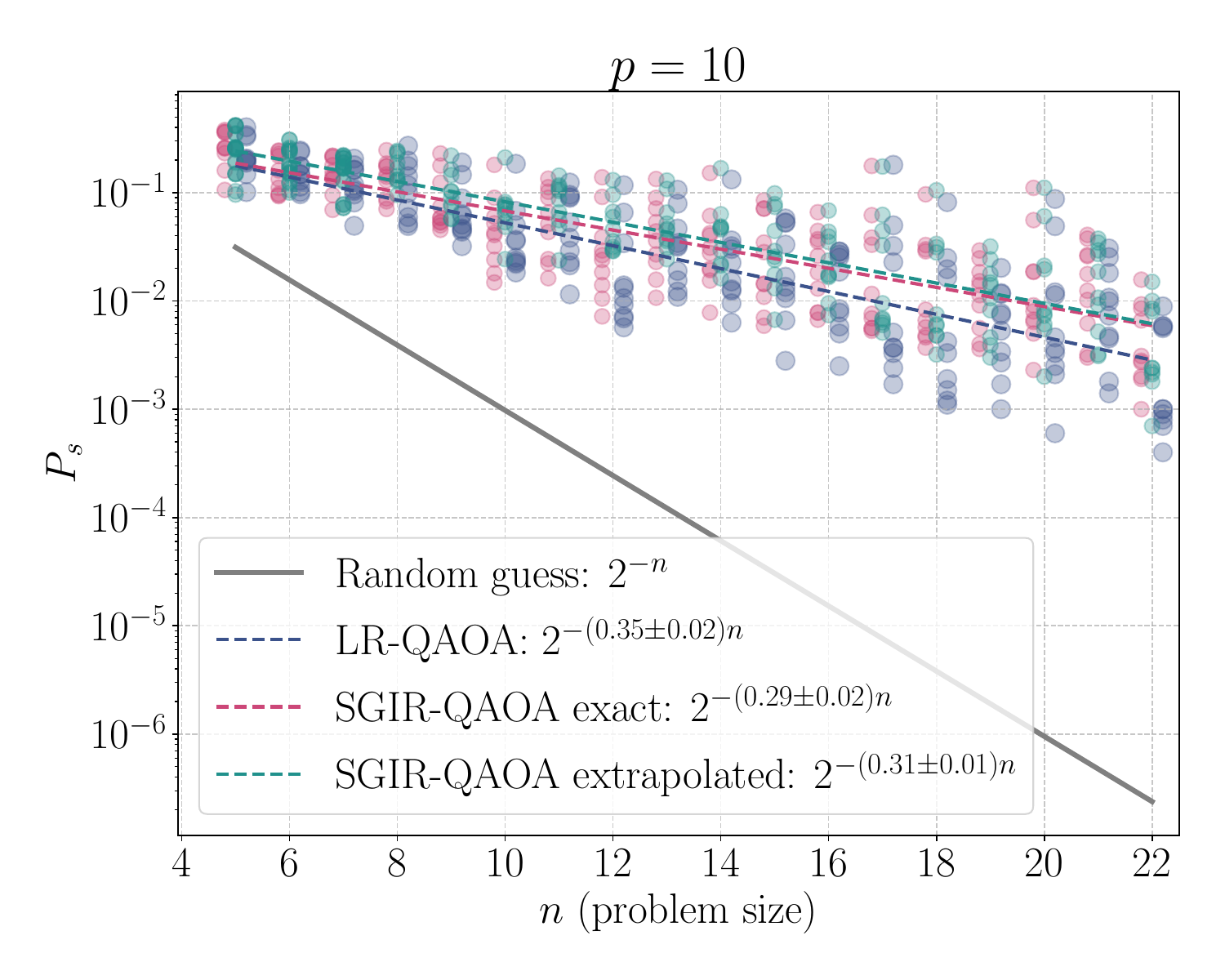}
\caption{\small{
Solving the MIS problem with ER graphs, probability of edges = 0.4, $\lambda=2000$. For extrapolated SGIR--QAOA, $n = 5-12$ is used to extrapolate $g_\text{min}$ which always occurs at $s=1$. At $s=0$ we take the gap $E_2-E_0=g_2 = 4$, as we know this analytically. For $s$ between 0 and 1 we take the average gap values from $n = 5-12$. At each $n$ the experiment is repeated 10 times with a different randomly generated instance. A log scale is used on the y-axis. The error in the fit for the exponential scaling coefficient is included in the legend.
}}
\label{plot:MIS_p04}
\end{figure}

We include here results on \emph{Erd\H{o}s--R\'enyi (ER) random graphs} with edge probability $p = 0.4$. In Figure \ref{plot:MIS_p04} results are shown using SGIR exact and the SGIR extrapolation technique. For the SGIR extrapolated, we use $g_\text{min}$ from $n = 6-12$ to make an exponential fit to extrapolate $g_{\text{min},s=1}$ for larger $n$.

\end{document}